# Neutron Scattering Studies of Pyrochlore Compound $Nd_2Mo_2O_7$ in Magnetic Field

Yukio Yasui[1], Satoshi Iikubo[1], Hiroshi Harashina[1], Taketomo Kageyama[1], Masafumi Ito[1] Masatoshi Sato[1] and Kazuhisa Kakurai[2]

[1]Department of Physics, Division of Material Science, Nagoya University,
Furo-cho, Chikusa-ku, Nagoya 464-8602

[2]Advanced Science Research Center, Japan Atomic Energy Research Institute, Tokai-mura, Naka-gun, Ibaraki, 319-1195

**Abstract**

Neutron diffraction studies have been carried out in the applied magnetic field $H$(//$[0\bar{1}1]$) on a single crystal of pyrochlore ferromagnet $Nd_2Mo_2O_7$, whose Hall resistivity($\rho_H$) has been reported to have quite unusual magnetic field ($H$)- and temperature ($T$)- dependences. The intensities of the observed magnetic reflections have been reproduced at 1.6 K as a function of $H$, by considering the change of the magnetic structure with $H$, where effects of the exchange fields at the Mo and Nd sites induced by the Mo-Mo and Mo-Nd exchange interactions and the single ion anisotropies of Mo- and Nd- moments are considered. From the $H$-dependent magnetic structure, the $H$-dependence of $\rho_H$ has been calculated by using the chiral order mechanism. By comparing the result with the $H$-dependence of the observed $\rho_H$, it is found that the chiral order mechanism does not work well in the present system.

## Introduction

Pyrochlore molybdate $R_2Mo_2O_7$ (R=Y and various rare earth elements) belongs to the pyrochlore series of compounds described by the general formula of $A_2B_2O_7$. The compounds have the face-centered cubic structure with the space group $Fd\bar{3}m$, in which A, B, O(1) and O(2) atoms fully occupy the sites 16d(A), 16c(B), 48f(O(1)) and 8b(O(2)), and the former two individually form three dimensional networks of corner-sharing tetrahedra.[1] The local principal axis of a corner site of a tetrahedron corresponds to the line which connects the site with the center of gravity of the tetrahedron. Because the single ion anisotropy of magnetic moments determined by the symmetry of local arrangement of surrounding atoms directs along the principal axis and because there are four different but crystallographically equivalent principal axes corresponding to the four corner sites, along [111] and other directions, non-collinear or non-coplanar magnetic structure is often expected.

In $R_2Mo_2O_7$, both the transport and the magnetic properties seem to exhibit systematic variation with the ionic radius of $R^{3+}$, that is, for the relatively larger elements R=Nd, Sm and Gd, the system exhibits the ferromagnetic transition, whereas for the smaller R=Tb-Lu and Y, it is insulating and exhibits the spin-glass-like behavior.[2-4] The spin-glass behavior is considered to originate from the geometrical frustration

$Nd_2Mo_2O_7$ is metallic and exhibits a ferromagnetic transition at a Curie temperature $T_C$=93 K. Neutron diffraction studies on a single crystal of $Nd_2Mo_2O_7$ were carried out and the magnetic structure in the temperature($T$) region of 4 K $\leq T \leq T_C$ has been proposed by the authors' group in refs. 5 and 6. In the $T$ region between 30 K and $T_C$, the ferromagnetic ordering at $T_C$=93 K is primarily associated with the Mo-moments, while the ordering of the Nd moments becomes significant below 30K. The net magnetization of the Nd-moments is antiparallel to that of the Mo-moments. The low temperature magnetic structure is non-collinear for both the Mo- and Nd-moments: The Mo-moments align along the direction nearly parallel to the [001] or other equivalent axes, but slightly tilted towards the local principal axes, while the Nd moments are along the directions almost parallel to their principal axes with their net magnetization being antiparallel to that of the Mo-moments. Because two of four Nd moments at corner sites of a tetrahedron direct inwards and the other two direct outwards of the tetrahedron, the magnetic structure is called "two-in and two-out" configuration.

Unusual behavior of the Hall resistivity ($\rho_H$) on single crystals of $Nd_2Mo_2O_7$ in the magnetic field ($\boldsymbol{H}$) along [111] or [001] direction has been reported by the authors' group.[7,8] In the $T$ region of 30 K<$T$<$T_C$, where the net magnetization is primarily from the Mo-moments, $H$-dependence of the Hall resistivity can be described by the well-known equation $\rho_H=R_0H+4\pi R_s M$ as for ordinary ferromagnets, where $R_0$ and $R_s$ are the ordinary and anomalous Hall coefficients, respectively, and $M$ is the total magnetization. In the temperature region of $T$<30 K, where the ordering of the Nd-moments is significant, the expression no more works and instead, another phenomenological equation $\rho_H=R_0H+4\pi R_s M_{Mo}+4\pi R_s'M_{Nd}$ has been found to describe the experimental results rather well, where $M_{Mo}$ and $M_{Nd}$, and $R_s$ and $R_s'$ are the net magnetizations and the anomalous Hall coefficient which correspond to the Mo- and Nd-moments, respectively. $R_s \times R_s'$ is found to be negative.(We defined the sign of $M_{Nd}$ to be negative if $\boldsymbol{M}_{Nd}$ is antiparallel to $\boldsymbol{M}_{Mo}$.) It is interesting that $R_s$ and $R_s'$ do not exhibit significant decrease as $T$ approaches zero, which is rather different from the behavior of the ordinary ferromagnets.[9,10] Although many theoretical works have been reported on the anomalous Hall effect, old theories which are based on the band picture[11] or on the mechanism of the conduction election scattering by localized moments[12], cannot describe this behavior. Effects of the Ti-doping into the Mo sites on the Hall resistivity have also been studied on single crystals of the present system, where it is found that the sign change of $R_s$ takes place with the small amount of the Ti-doping with the sign of $R_s$' unchanged.[13] This indicates that two components are necessary to describe the behavior of $\rho_H$ of $Nd_2Mo_2O_7$.

A new proposal has been made that the ordering of the spin chirality($\chi$) contributes to the Hall resistivity, where $\chi$ is locally defined as $\chi \equiv S_1 \cdot S_2 \times S_3$ for three spins $S_1$, $S_2$ and $S_3$.[14] In this theory, the anomalous Hall conductivity $\sigma_H(=\rho_H/\rho^2)$ is proportional to the fictitious magnetic flux($\Phi$) induced by the spin chirality $\chi$, where $\rho$ is the electrical resistivity. Taguchi *et al.* proposed that the spin chirality or the fictitious magnetic flux of the Mo-moments is playing an important role in determining the unusual behavior of the Hall resistivity.[15] However, authors' group has reported that the idea does not seem to consistently explain the observed behavior.[5,6,13] It is important for clarifying the relationship between the spin chiral order and the anomalous Hall resistivity, to study the detailed magnetic structure of $Nd_2Mo_2O_7$, particularly to study the magnetic structure in the applied magnetic field.

In the present work, neutron scattering studies have been carried out on a single crystal of $Nd_2Mo_2O_7$ in the applied magnetic field $H(//[0\bar{1}1])$ up to 5.7 T. The $H$-dependence of the magnetic structure is naturally understood by considering the Mo-Mo and Mo-Nd exchange fields and the single ion anisotropies of the Mo- and Nd-moments. Then, the spin chirality and the fictitious magnetic flux have been calculated as a function of $H$ at $T$=1.6 K and as a function of $T$ at $H$=0. Comparing the Hall resistivity $\rho_H$ deduced from the fictitious magnetic flux with the experimentally observed data, we have found that the unusual behavior of the Hall resistivity of $Nd_2Mo_2O_7$ cannot consistently be understood by the chiral order mechanism in the present system.

**Experiments**

A single crystal of $Nd_2Mo_2O_7$ with a volume of ~0.3 cm$^3$ was grown by a floating zone (FZ) method in Ar atmosphere. The magnetization and the Hall resistivity were measured by using edge parts of the crystal. The crystal planes of the samples were determined by observing the X-ray or neutron diffraction lines. The magnetization $M$ and the Hall resistivity $\rho_H$ were measured as a function of $H$ by using a SQUID magnetometer, as described in refs.7 and 8.

Neutron measurements were carried out by using the triple axis spectrometer HQR(T1-1) installed at the thermal guide of JRR-3M of JAERI in Tokai, where the double axis condition was adopted. The crystal was oriented with the $[0\bar{1}1]$ direction vertical, where the [011] and [100] axes are in the scattering plane. The magnetic field was applied by using a superconducting magnet along the vertical or $[0\bar{1}1]$ direction. The 002 reflections of Pyrolytic graphite (PG) were used for both the monochromator and analyzer. Horizontal collimations were 12'(effective)-20'-60' and the neutron wavelength was 2.460 Å. A PG filter was put in front of the sample to suppress the higher-order contamination.

**Experimental Results and Discussion**

In the fields $H$=0 and 5 T, the neutron elastic scattering intensity has been measured on the single crystal of $Nd_2Mo_2O_7$ at various $\boldsymbol{Q}$-points in the reciprocal space including those with half integer values of $h$ and $k$(=$l$) at 113 K(>$T_C$) and 1.6 K(<$T_C$). No reflection has been detected at $\boldsymbol{Q}$-points except the nuclear Bragg points. The scattering intensities at several nuclear Bragg points increases with decreasing $T$ through the Curie temperature. The intensities also depend on the values of the external magnetic field $H$ applied along //[$0\bar{1}1$]. Examples of the observed profiles of several reflections are shown in Figs. 1(a)-1(d), where the data at 113 K(>$T_C$) are the contribution from the nuclear reflection. In the present experiment, we have not observed the 002 nuclear reflection and all the nuclear reflections can be explained by the undistorted pyrochlore structure (space group F$d\bar{3}m$). The observation of the reflection in the previous measurements[5,6)] is considered to be due to the imperfect removal of the high-order contamination. (The same crystal was used in the present and previous studies.) The profile widths of all reflections observed at 1.6 K under the magnetic field are equal to those of the instrumental resolution within the experimental error bars.

The magnetic scattering intensities $I_{mag}$ of $Nd_2Mo_2O_7$ of the 111, 200, 022, 311 and 400 reflections are shown at 1.6K against the magnetic field $H$($\boldsymbol{H}$//[$0\bar{1}1$]) in Figs. 2(a)-2(e), respectively, where the data are taken by scanning $H$ stepwise up to 5.7 T after zero field cooling and then to zero. Hysteretic behavior of the $I_{mag}$ –$H$ curves has been observed in the region of $H$ ≤1.5 T. It is due to the hysteretic domain motion. $I_{mag}$ was obtained by taking the differences between the integrated intensities below and above $T_C$, where the absorption- and Lorentz factor-corrections were considered. The extinction effect has been estimated by plotting the observed values of the absolute structure factors $|F(\boldsymbol{Q})|^2$ of the nuclear scattering determined at 113K against those estimated by adopting the positional parameter $z$ =0.333 for 48f(O(1)) sites. Details of the extinction correction have already been reported in ref. 5. Effects of the error of the extinction correction do not bring about essential changes of the results of the present study, because only the data at the Bragg points where the extinction effect is not serious (or scattering intensity is relatively weak) are mainly considered in the magnetic structure analysis.

The system has the f.c.c. unit cell and the crystal structure consists of two networks individually formed by corner sharing tetrahedra of $Mo_4$ or $Nd_4$ as mentioned above. In this f.c.c. cell, each one of four primitive cells has a structure unit consisting of a $Mo_4$-tetrahedron and a $Nd_4$-one. In the measurements at 1.6 K in the zero- and finite magnetic fields, the magnetic reflections have been observed only at $\boldsymbol{Q}$-points with even $h$ and $k$ or odd $h$ and $k$, where $h$ and $k$ are defined with respect to the f.c.c. unit cell. This result indicates that the magnetic primitive cell of $Nd_2Mo_2O_7$ has the same size as the chemical primitive cell. Then, to determine the magnetic structure of the present system, we are just required to find the moment arrangements within each one of $Mo_4$- and $Nd_4$-tetrahedra (or within a magnetic primitive cell ).

First, we present the data taken at 1.6 K in zero magnetic field. The ordering patterns which can reproduce the intensity distribution of $I_{mag}$ are shown for the $Mo_4$- and $Nd_4$- tetrahedra in Figs. 3(a) and 3(b), respectively, where thick arrows indicate the directions of the magnetic moments and G indicates the center of gravity of the tetrahedron. The net magnetizations $M_{Nd}$ and $M_{Mo}$ are along the [001] or other equivalent directions (We have chosen this to be the [001] direction ($z$ axis) in Figs. 3(a) and 3(b).). The intensity distribution of $I_{mag}$ calculated for the pattern is listed in Table I. In the calculation, we used the reported values of the magnetic form factors for $Nd^{3+}$ moments,[16)] and for $Mo^{4+}$, values experimentally determined for $MoF_3$ were used.[17)] The distribution probability of the ferromagnetic domains among the three equivalent magnetization directions, [001], [010] and [100] is optimized.

The Mo-moments align along the direction nearly parallel to the $z$ axis but have the slight tilting by the angle $\alpha = 4(\pm 4)°$ from the $z$ axis towards the corresponding local principal axes. The Nd-moments are along the directions nearly parallel to their principal axes with their $z$-component being antiparallel to the $z$ axis, but the possibility of the tilting by $\pm 3°$ from their principal axes cannot be excluded. The results indicate that there exists the relatively strong axial anisotropy at the Nd sites. The obtained magnetic ordering pattern indicates that the Mo-Mo and Mo-Nd interactions are ferromagnetic and antiferromagnetic, respectively. The values of the ordered atomic moments of Mo and Nd, $\mu_{Mo}$ and $\mu_{Nd}$, respectively, are estimated by comparing the magnetic scattering intensities with the nuclear ones, to be about $1.30(\pm 0.05)\mu_B$ and $1.71(\pm 0.05)\mu_B$. The proposed ordering pattern does not contradict the pattern at 4.0 K reported in refs. 5 and 6. In the temperature region of $T \leq 30$ K, it is considered that $\mu_{Mo}$ is almost $T$-independent and $\mu_{Nd}$ increases with decreasing $T$. Considering the local arrangement of the Nd-moments which surround a Mo site, we expect that the ordered Mo-moments are tilted by the Mo-Nd antiferromagnetic exchange interaction towards the direction opposite to the local principal one. Then, with decreasing $T$, the values of $\alpha$ is expected to decrease.

Now, the ordering pattern of $Nd_2Mo_2O_7$ in the external magnetic field $H(//[0\bar{1}1])$ is discussed. The direction of the ordered Mo-moment at the $i$-th site, $\boldsymbol{\mu}_{Mo,i}$ (The indices $i(=1-4)$ indicate the four corner sites within a tetrahedron as shown in Figs. 5(a) and 5(b).) can be determined by the relation

$$\boldsymbol{\mu}_{Mo,i} // (\boldsymbol{H} + \boldsymbol{H}_{Mo \leftarrow Mo} + \boldsymbol{H}_{Mo \leftarrow Nd,i} + \boldsymbol{H}_{ani}^{Mo}) \qquad (1)$$

where $\boldsymbol{H}_{Mo \leftarrow Mo}$ and $\boldsymbol{H}_{Mo \leftarrow Nd,i}$ are the exchange fields from the surrounding Mo- and Nd-moments, respectively. $\boldsymbol{H}_{ani}^{Mo}$ is the effective magnetic field which describes the effect of the single ion anisotropy at the Mo-sites. The absolute value of $\boldsymbol{\mu}_{Mo,i}$ at 1.6 K is equal to be 1.30 $\mu_B$ and is considered to be $H$-independent. The absolute value of $\boldsymbol{H}_{Mo \leftarrow Mo}$ is estimated from the value of the Curie temperature $T_C$=93 K, which is primarily due to the ordering of the Mo-moments, is to be $\sim$67 T. The direction of the ordered Nd-moment at the $j$-th site($j$=1-4), $\boldsymbol{\mu}_{Nd,j}$ is fixed along the [111] or other equivalent directions, by considering that the Nd-moments are in the doublet levels and have the strong Ising-like anisotropy, where the absolute value of the ordered Nd-moment, $\mu_{Nd,j}$ can be described by the equations,

$$\mu_{Nd,j}(T) = \mu_{Nd,j}(T=0) \cdot \tanh\left( \frac{\mu_{Nd,j}(T=0) \cdot (\boldsymbol{H}_{Nd \leftarrow Mo,j} - \boldsymbol{H})}{k_B T} \right) \qquad (2)$$

where $\boldsymbol{H}_{Nd \leftarrow Mo,j}$ is the exchange field at the $j$-th Nd site from the surrounding Mo-moments. Because the Weiss temperatures of $Nd_2Zr_2O_7$ and $Nd_2GaSbO_7$, which have the Nd-Nd atomic distances similar to that of $Nd_2Mo_2O_7$, is smaller than 1 K[18,19)], we expect that the interaction among the Nd-moments in $Nd_2Mo_2O_7$ is considered to be negligible. The absolute value of the ordered moment of Nd, $\mu_{Nd,j}$ at $T$=1.6

K is equal to be 1.71 $\mu_B$. $\boldsymbol{H}_{Mo \leftarrow Nd,i}$ and $\boldsymbol{H}_{Nd \leftarrow Mo,j}$ can be described by the equations as,

$$\boldsymbol{H}_{Mo \leftarrow Nd,i} = J_{Mo-Nd} \sum_j \boldsymbol{\mu}_{Nd,j}(T) \qquad (3)$$

$$\boldsymbol{H}_{Nd \leftarrow Mo,j} = J_{Mo-Nd} \sum_i \boldsymbol{\mu}_{Mo,i}(T) \qquad (4)$$

where $J_{Mo-Nd}$ is the coupling constant between the Mo- and Nd-moments and the summations are taken over the nearest neighbor sites. The dipole-dipole interaction between the Mo- and Nd-moments is of the order of 0.1 K and can be neglected here. Schematic figures of the directions of these fields at the Mo- and Nd-sites are shown in Figs. 4(a) and 4(b), respectively. We have chosen the values of $\mu_{Nd,j}(T=0)$, $J_{Mo-Nd}$ and $H_{ani}^{Mo}$ to reproduce the $I_{mag}$ data obtained at the Bragg points listed in Table I and to reproduce the $H$-dependence of $I_{mag}$ shown in Figs. 2(a)-2(e), where $I_{mag}$ is scaled by the nuclear scattering intensities: $\mu_{Nd,j}(T=0)=1.93$ $\mu_B$, $J_{Mo-Nd}=-0.39$ T/$\mu_B$ and $H_{ani}^{Mo}=7.6$ T.

Once these values are determined, we can derive the magnetic structure of the present system for arbitrary $\boldsymbol{H}$ and $T$. The obtained ordering patterns at 1.6 K in the magnetic field $H \to 0$ and $H=5.7$ T are shown in Figs. 5(a) and 5(b), respectively, where $H \to 0$ indicates that the magnetic field is extrapolated to zero from the relatively high magnetic field, and that the direction of the ferromagnetic component is uniquely given. The $\mu_{Nd,j}$ values ($j$=1-4) are shown against the magnetic field $H$ at 1.6 K in Fig. 5(c). $I_{mag}$ calculated for these patterns are shown against $H$ in Figs. 2(a)-2(e) by solid lines. In the calculation, the intensity average has been taken for crystallographically equivalent Bragg points in the scattering plane, the plane perpendicular to $H$, to consider the degrees of freedom of the domain structure about the direction of $\boldsymbol{H}$. The Mo-moments align along the direction nearly parallel to $\boldsymbol{H}$(//$[0\bar{1}1]$), but they are tilted from $[0\bar{1}1]$ direction by the $\boldsymbol{H}_{ani}^{Mo}$ and $\boldsymbol{H}_{Mo \leftarrow Nd,i}$, with their tilting directions and angles depending on $H$ through $H$-dependence of the magnitude of $\mu_{Nd,j}(T)$. The directions of $\mu_{Nd,j}$ ($j$=1 and 3) are nearly $H$-independent as shown in Figs. 5(a) and 5(b), because their directions (parallel to the corresponding principal axes) are perpendicular to $\boldsymbol{H}$ and nearly perpendicular to $\boldsymbol{H}_{Nd \leftarrow Mo,j}$. The values of the ordered Nd-moment $\mu_{Nd,j}$ ($j$=2 and 4) turn out, as shown in Fig. 5(c), to become zero at $H$=3.0 T, which can be considered to correspond to $\boldsymbol{H}_{Nd \leftarrow Mo,j}$.

In Fig. 6, the $M$-$H$ curves of a single crystal of Nd$_2$Mo$_2$O$_7$ taken at various temperatures after the zero field cooling under the field $\boldsymbol{H}$//$[0\bar{1}1]$ (or equivalently $\boldsymbol{H}$//[110]), is shown together with those calculated at 1.6 K and 3.0 K for the magnetic structure proposed above. We have found that the calculated data is quantitatively consistent with the experimentally observed data, though the difference is observed in the $H$-region of $H \leq 1$ T, which arises from the fact that the calculated results are obtained for the magnetic state with the ferromagnetic component of all domains being aligned in the direction of $H$.

We have derived the magnetic structure of Nd$_2$Mo$_2$O$_7$ in the magnetic field $\boldsymbol{H}$(//$[0\bar{1}1]$) by using the relations (1)-(4) and by using the values of $\mu_{Nd,j}(T=0)$, $J_{Mo-Nd}$ and $H_{ani}^{Mo}$ deduced from the experimental results. We have also derived the magnetic structures at 1.6 K under the conditions $\boldsymbol{H}$//[001] and [111] for the same values of $\mu_{Nd,j}(T=0)$, $J_{Mo-Nd}$ and $H_{ani}^{Mo}$ as those used above. We have further tried to determine the magnetic structures by using $H_{ani}^{Mo}=0$ instead of $H_{ani}^{Mo}=7.6$ T, with keeping $\mu_{Nd,j}(T=0)$ and $J_{Mo-Nd}$ unchanged, though the degree of agreement between the calculated and the observed $I_{mag}$ distributions for $H_{ani}^{Mo}=0$ is slightly worse than that for $H_{ani}^{Mo}=7.6$ T.

In the above calculations, the rather strong Ising-like anisotropy of the Nd moments is essential to reproduce the observed $H$-dependence of the magnetic scattering intensities, where the ordered Nd moment at $T=0$ $\mu_{Nd,j}(T=0)$ is found to be smaller than the moment value $g\mu_B J$ of the free Nd$^{3+}$ ions ($J=9/2$). To understand this results, we consider a doublet with the expectation values of the angular momentum component along the principal axis, $\pm |J'|$ ($J'<J$), where the strong anisotropy can be expected if the energies of the lowest excited states are much larger than $\mu_{Nd,j}(T=0) \cdot H$ and $k_B T$. The components of $J$ perpendicular to the principal axis are considered not to have the ordering.

In Figs. 7(a)-7(c), the Hall resistivities $\rho_H$ of a single crystal of Nd$_2$Mo$_2$O$_7$ are shown at various temperatures against the magnetic fields $H$ applied along the [110], [001] and [111] directions, respectively,[7,8,13] The authors' group reported that the $H$- and $T$- dependences of $\rho_H$ of Nd$_2$Mo$_2$O$_7$ are quite unusual: The behaviors of $\rho_H$ are markedly different between the $T$-regions above and below ~30 K. Because the total magnetization is determined by the sum of the net magnetizations $M_{Mo}$ and $M_{Nd}$ which have opposite signs in the weak field region, the enhancement of $\rho_H$ cannot be reproduced by the expression $\rho_H = R_0 H + 4\pi R_s M$. Instead, the phenomenological expression $\rho_H = R_0 H + 4\pi R_s M_{Mo} + 4\pi R_s' M_{Nd}$ has been, as stated above, found to reproduce the results rather well, indicating that the perturbative treatment is valid. However, any of the old theories cannot describe the $T$-dependence of the Hall resistivity $\rho_H$ of Nd$_2$Mo$_2$O$_7$.

Ohgushi et al.[13] and Taguchi et al.[14] have proposed that the ordering of the spin chirality $\chi$ contributes to the Hall resistivity. According to this mechanism, the fictitious magnetic flux $\phi$ induced by the spin chirality $\chi$ acts on the electron motion in the same way as the real magnetic field, where $\rho_H$ cannot be treated by the perturbation of the spin-orbit interaction. The absolute value of the fictitious flux, $|\phi_i|$ ($i=1$~$4$) is proportional to $\chi \equiv S_j \cdot S_k \times S_l$, where $\{j, k, l\}$ being the $i$-th set of different numbers chosen in a cyclic way from 1-4 to specify the set of spins within a tetrahedron, and its direction is defined as shown in Fig. 8 for $\phi_1$ for example. The total fictitious magnetic flux $\Phi$ of the system is vector sum of $\phi_1$~$\phi_4$. The anomalous Hall conductivity $\sigma_H (= \rho_H/\rho^2)$ is proportional to the parallel component of $\Phi$ to the external magnetic field.

The fictitious magnetic flux of the Mo- and Nd-moments of Nd$_2$Mo$_2$O$_7$, $\Phi_{Mo}$ and $\Phi_{Nd}$, respectively, has been calculated at 1.6 K as a function of $H$ by using the $H$-dependent magnetic structure derived above. Their parallel components to the external magnetic field $H$, $\Phi_{Mo}^{//}$ and $\Phi_{Nd}^{//}$, are shown in Figs. 9(a)-9(c), for the cases $H$//[110], [001] and [111], respectively. In the figures, the results obtained for both $H_{ani}^{Mo}=7.6$ T and $H_{ani}^{Mo}=0$ are plotted (Note that $\Phi_{Nd}^{//}$ hardly depend on the value of $H_{ani}^{Mo}$.). Because the electrical resistivity $\rho$ of Nd$_2$Mo$_2$O$_7$ below ~60 K is almost $T$- and $H$-independent[13,19] in the $H$-region of $H \leq 7$ T, the Hall conductivity $\sigma_H (=\rho_H/\rho^2)$ is nearly proportional to the Hall resistivity $\rho_H$. It should be also noted that $\rho_H$ does not significantly depend on $T$ below 3 K.[13] Then, according to the chiral order mechanism, the $H$-dependences of $\Phi_{Mo}^{//}$ and $\Phi_{Nd}^{//}$, at 1.6 K shown in Figs 9(a)-9(c) should agree with those of the Hall resistivity $\rho_H$ observed at 3 K, which are shown in Figs. 7(a)-7(c), respectively. However, we cannot find the agreement in the figures: For $H$ //[001], for example, the observed $\rho_H$ decreases monotonically with increasing $H$, while the calculated $\Phi_{Mo}^{//}$ for $H_{ani}^{Mo}=7.6$ T increases with increasing $H$ and $\Phi_{Mo}^{//}$ for $H_{ani}^{Mo}=0$ does not exhibits the strong decreasing behavior. ( We note here that the ordinary Hall resistivity is not significantly large and its consideration does not change the above arguments.) As stated above, the observed $\rho_H$-$H$ curves can phenomenologically be described by using two degrees of freedom which contribute to the anomalous Hall resistivity $\rho_H$[13]: One is proportional to $M_{Mo}$ and the other is proportional to $M_{Nd}$. However, it is difficult to explain the observed $\rho_H$ data by the linear combination of the components of $\Phi_{Mo}^{//}$ and $\Phi_{Nd}^{//}$,

Now, the $T$-dependence of the fictitious magnetic flux of the Mo- and Nd-moments, $\Phi_{Mo}^{//}$ and $\Phi_{Nd}^{//}$ is discussed. We calculate the fictitious flux by assuming that even at $H=0$, both of $\Phi_{Mo}$ and $\Phi_{Nd}$, are parallel to the [001], or the ferromagnetic domains are aligned. Figure 10(a) shows their absolute values at $H=0$ divided by the corresponding value at 1.6 K. In the calculation, the ordered Nd-moments $\mu_{Nd,j}(T)$ has been estimated from the temperature dependence of the observed magnetic scattering intensity $I_{mag}$ of the 111 reflection shown in the inset of Fig. 10(a). For $H_{ani}^{Mo}=7.6$ T, $\Phi_{Mo}^{//}$ decreases with decreasing $T$ in contrast to that at $H_{ani}^{Mo}=0$. Figure 10(b) shows the $T$-dependence of the Hall resistivity $\rho_H$ of Nd$_2$Mo$_2$O$_7$ taken under the relatively weak magnetic field $H=0.5$ T along the [001] direction. Because at $H=0.5$ T, the magnetic domains are aligned but the magnetic structure is not significantly changed from that at $H=0$, the observed $\rho_H$ can be compared with the above calculated results at $H=0$. We find, however, that the $T$-dependences of $\Phi_{Mo}^{//}$ and $\Phi_{Nd}^{//}$, are quite different from that of the observed Hall resistivity as shown in Figs. 10(a) and 10(b), irrespective of the choice of the value of $H_{ani}^{Mo}$. Thus, the $T$- and $H$-dependences of the fictitious magnetic flux deduced from the proposed magnetic structure do not agree with that of the observed Hall resistivity $\rho_H$ of Nd$_2$Mo$_2$O$_7$, indicating that unusual behavior of $\rho_H$ is not explained only by the chiral order mechanism.

In the above calculations of the fictitious magnetic flux, we have estimated the ordered Nd-moments $\mu_{Nd,j}$ by using the mean field approximation. The problem might remain whether the spin chirality or the fictitious magnetic flux defined by the local arrangement of moments could properly be obtained from this $\mu_{Nd,j}$. For example, even in the case where $H_{ani}^{Mo}=0$ and $\mu_{Nd,j}=0$, non-zero $\Phi_{Mo}^{//}$ is induced through the exchange interaction of Mo- and Nd-moments, if only the Mo-moments are ferromagnetically ordered. However, the value of $\Phi_{Mo}^{//}$ induced in the situation is too small(by a factor of ~1/50) to explain the $\rho_H$ value observed in the region of $T>40$ K. Then, it is very difficult to expect significant effects of the local fluctuation of the chirality to the behavior of $\rho_H$. The observed behavior of $\rho_H$ of the present system cannot be understood by considering the chiral order as the main mechanism of the Hall transport.

In summary, we have shown the neutron scattering data taken for a single crystal of Nd$_2$Mo$_2$O$_7$ in the $T$-region from 1.6 K to 150 K ($>T_C$) and under the magnetic field up to 5.7 T applied along the [0$\bar{1}$1] direction. The magnetic structure in the applied magnetic field $0 \leq H \leq 5.7$ T is derived by estimating the exchange coupling constant between the Mo- and Nd-moments and by considering the single ion anisotropies of each Mo- and Nd-moments. From the proposed magnetic structure, the fictitious magnetic flux of the Mo- and Nd-moments has been calculated as functions of $T$ and $H$ and compared with the observed data of the Hall resistivity $\rho_H$. The results indicate that the chiral order mechanism is not, at least, playing a main role in determining the observed behavior of $\rho_H$ in the present system.

Acknowledgements

The authors thank Prof. Nagaosa of the University of Tokyo for stimulating discussion. They also thank the Neutron Scattering Laboratory of the Institute for Solid State Physics(NSL-ISSP) and Research Center for Nuclear Science and Technology(RCNST) of the University of Tokyo for the use of the instrument within the national user's program.

Figure captions

Fig. 1. Profiles of the ω-scans for 111, 200, 022 400 reflections taken under the magnetic field $H(//[0\bar{1}1])$ are shown in (a)-(d), respectively. The values of $H$ and $T$ are shown in the figures.

Fig. 2. Magnetic field $H$- ($H//[0\bar{1}1]$) dependence of the magnetic scattering intensities of $Nd_2Mo_2O_7$ at 1.6 K is shown for 111, 200, 022, 311 and 400 reflections in (a)-(e), respectively, where the data were taken after zero field cooling by scanning $H$ stepwise up to 5.7 T and then down to zero. Solid lines show the calculated magnetic scattering intensities for the magnetic ordering pattern derived here. See text for details.

Fig. 3. The magnetic ordering patterns which can explain the observed magnetic scattering intensities of $Nd_2Mo_2O_7$ taken at $T$=1.6 K and $H$=0, are shown for the $Mo_4$- and $Nd_4$-tetrahedra in (a) and (b), respectively, where thick arrows indicate the directions of the magnetic moments. G indicates the center of gravity of the tetrahedra.

Fig. 4. Schematic figure of the magnetic fields which act on the (a)Mo- and (b) Nd-moments. $H$ is the external magnetic field along $[0\bar{1}1]$ direction. $H_{ani}^{Mo}$ and $H_{ani}^{Nd}$ are the effective magnetic fields derived from the single-ion anisotropy of the Mo- and Nd- moments, respectively. $H_{Mo\leftarrow Mo}$ and $H_{Mo\leftarrow Nd,j}$ are the exchange fields at the $i$-th Mo site within a tetrahedron from the surrounding Mo- and Nd- moments, respectively. $H_{Mo\leftarrow Nd,j}$ is the exchange field at the $j$-th Nd site within a tetrahedron from the surrounding Mo-moments.

Fig. 5. Magnetic ordering patterns of the $Mo_4$- and $Nd_4$-tetrahedra, which can explain the observed magnetic scattering intensities of $Nd_2Mo_2O_7$ taken at 1.6 K under the magnetic fields (a) $H \rightarrow 0$ and (b) 5.7 T, respectively. Thick and thin arrows indicate the directions of the magnetic moments and the external magnetic field $H(//[0\bar{1}1])$, respectively. The absolute value of the ordered Nd-moments, $\mu_{Nd,j}$ ($j$=1-4) are shown at 1.6 K against the magnetic field $H(//[0\bar{1}1])$ in (c). See text for details.

Fig. 6. Magnetization $M$ of $Nd_2Mo_2O_7$ is shown at various temperatures against the magnetic field $H$ applied along $[0\bar{1}1]$, where dotted and solid lines show the results of the calculations at 1.6 K and 3.0 K, respectively, for the magnetic ordering pattern derived here( See Figs. 5(a)-5(c).).

Fig. 7. Hall resistivity $\rho_H$ of $Nd_2Mo_2O_7$ is shown at various temperatures against the magnetic field $H$ applied along (a) [110], (b) [001] and (c) [111], respectively. The data in (b) and (c) were reported previously by authors' group.[7,8,13]

Fig. 8. Schematic figure of the directions of the magnetic moments and the fictitious magnetic flux within a tetrahedron. The direction of the fictitious flux $\phi$ is perpendicular to the triangle found by the sites of three spins and its absolute value $|\phi|$ is proportional to their spin chirality $\chi$. The total fictitious magnetic flux $\Phi$ is the sum of the four vectors of the fictitious flux, $\phi$ determined by the three magnetic moments periodically chosen among the four moments within a tetrahedron.

Fig. 9. Calculated fictitious magnetic flux of the Mo- and Nd-moments along the external magnetic field, $\Phi_{Mo}{}^{//}$ and $\Phi_{Nd}{}^{//}$, respectively, is shown at 1.6 K against the magnetic fields $H$ applied along (a) [110], (b) [001] and (c) [111], where solid and long dashed lines are for the Mo-moments in the case of $H_{ani}^{Mo}$=7.6 T and $H_{ani}^{Mo}$=0, respectively, and the dashed line is

| | for the Nd-moments. |
|---|---|
| Fig. 10. | (a) Temperature dependence of the absolute values of the calculated fictitious magnetic flux of the Mo- and Nd-moments, $\Phi_{Mo}$ and $\Phi_{Nd}$, respectively, is shown at $H$=0 (Both are parallel to [001]). The values are scaled by the data at 1.6 K. Inset shows the temperature dependence of the observed magnetic scattering intensity of the 111 reflection taken for a single crystal of $Nd_2Mo_2O_7$. (b) Temperature dependence of the Hall resistivity $\rho_H$ of $Nd_2Mo_2O_7$ taken under the magnetic field $H$=0.5 T along the [001] direction. Broken lines are the guides for the eye. |

Table. I. Integrated intensities of the magnetic scattering taken at 1.6 K under the zero external field are compared with those of the model calculations at several reflection points.

| | $T=1.6$K, $H=0$ | |
|---|---|---|
| hkl | $I_{obs}$ | $I_{calc}^{*}$ |
| 111 | 339.2 ± 22.4 | 318.3 |
| 200 | 116.7 ± 6.0 | 112.0 |
| 022 | 37.9 ± 3.9 | 37.0 |
| 311 | 58.2 ± 5.9 | 63.9 |
| 400 | 10.8 ± 4.1 | 10.2 |
| 133 | 36.2 ± 21.1 | 44.6 |
| 422 | 23.7 ± 10.4 | 26.8 |
| 511 | 32.3 ± 10.5 | 27.2 |
| 600 | 21.4 ± 1.5 | 21.4 |

$^{*}$for $\mu_{Mo}=1.30\mu_{B}$, $\mu_{Nd}=1.71\mu_{B}$

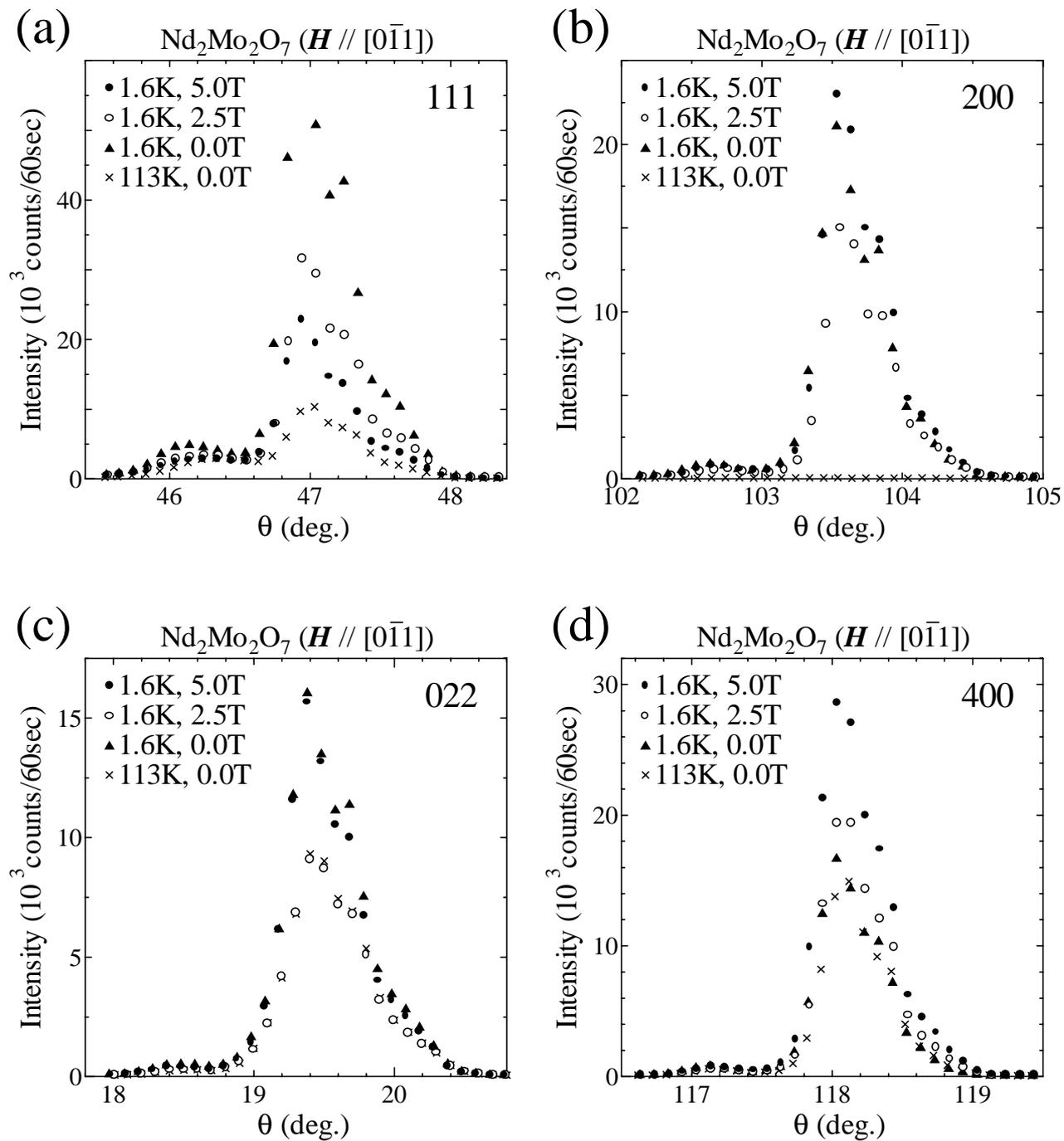

Fig. 1

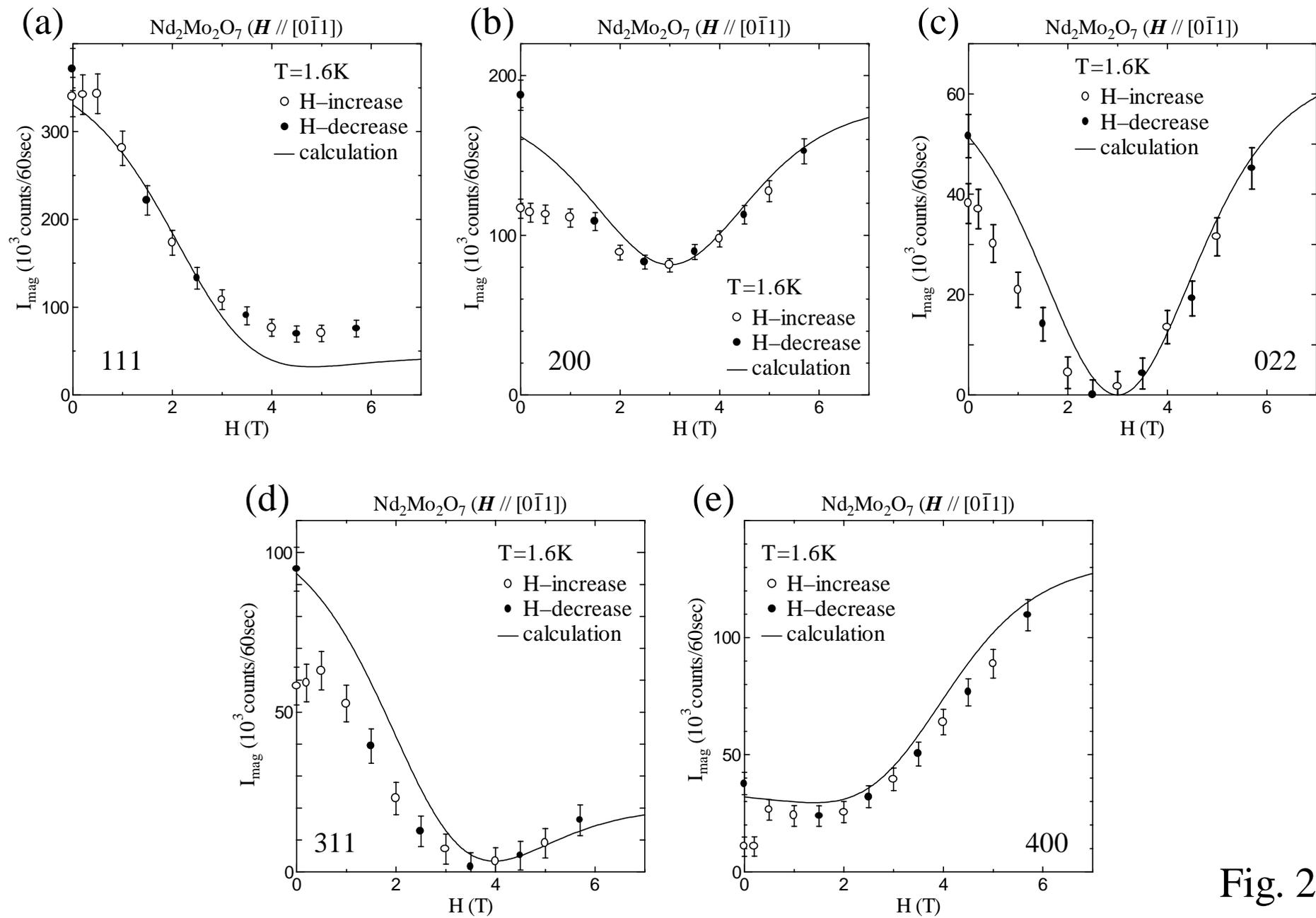

Fig. 2

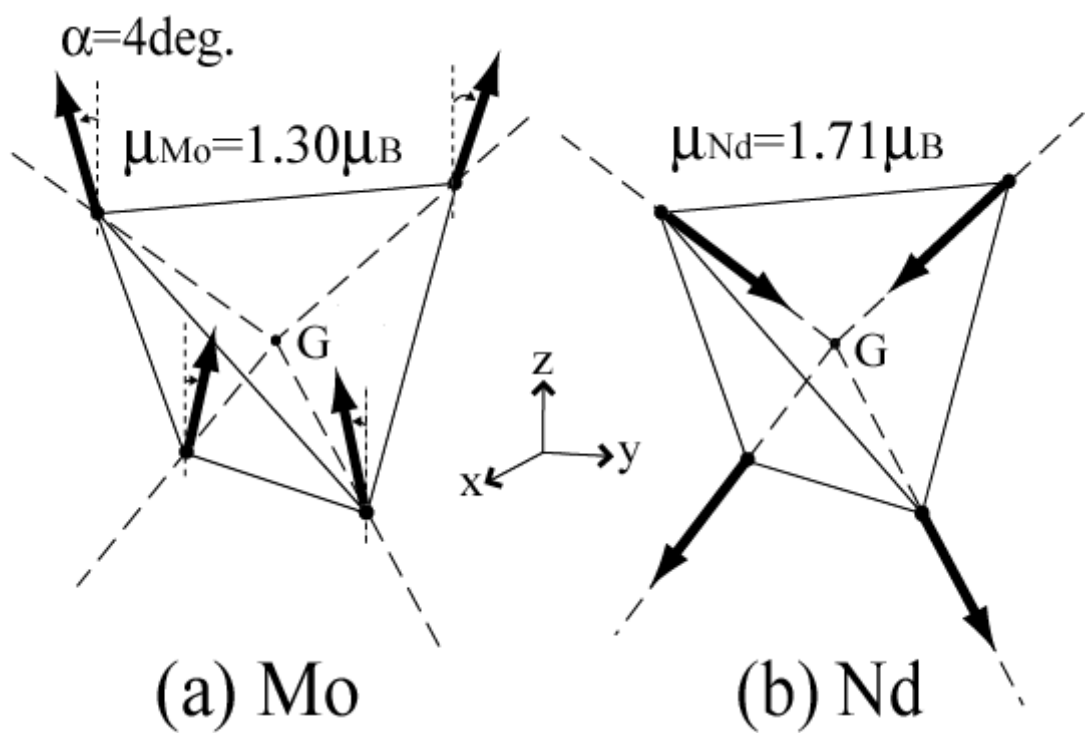

Fig. 3

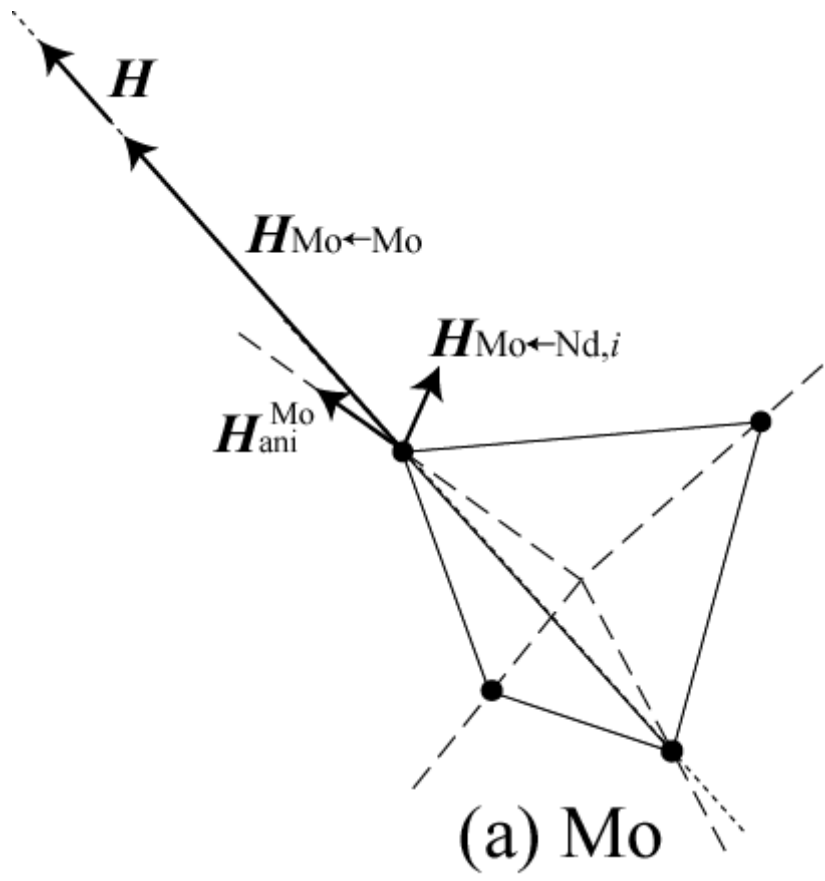

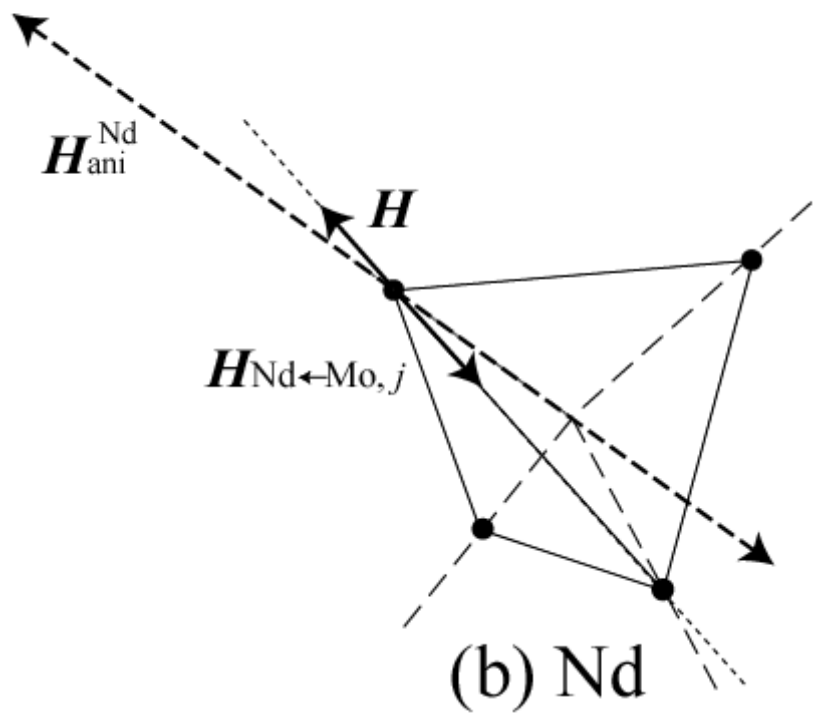

Fig. 4

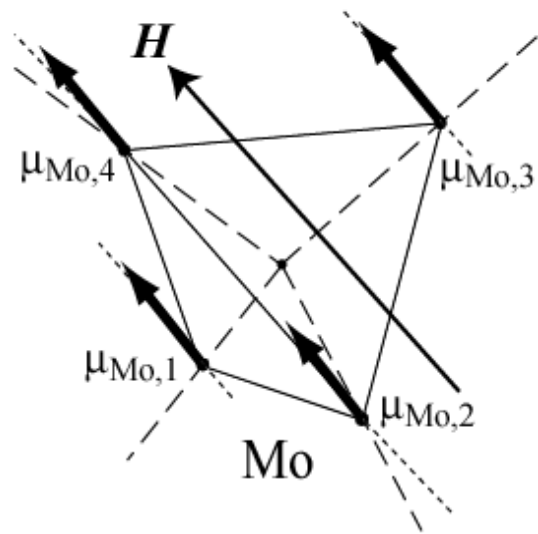
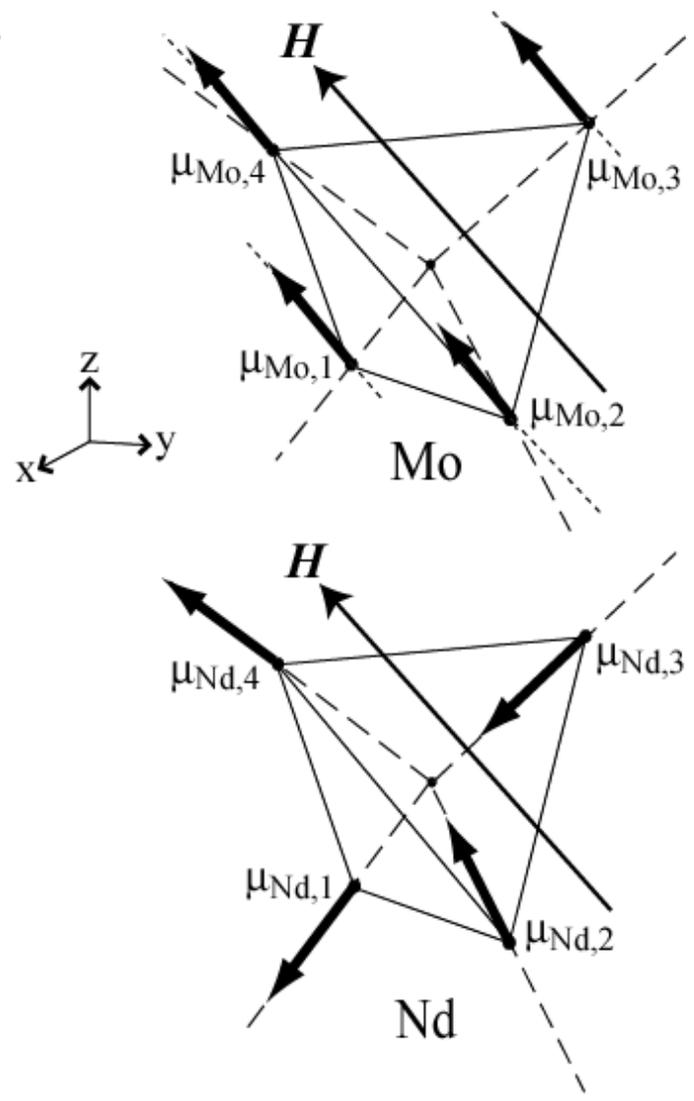
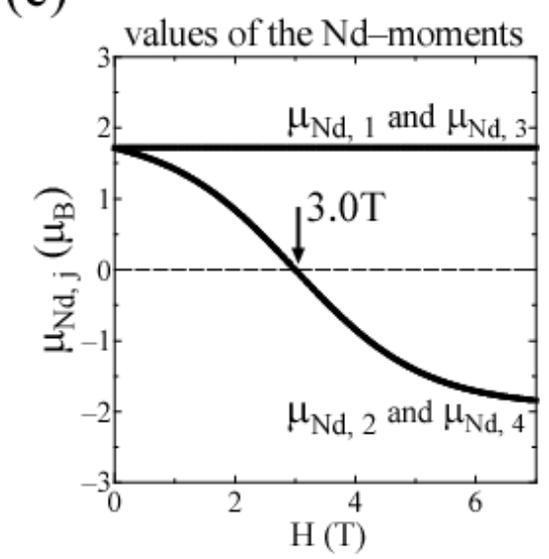

Fig. 5

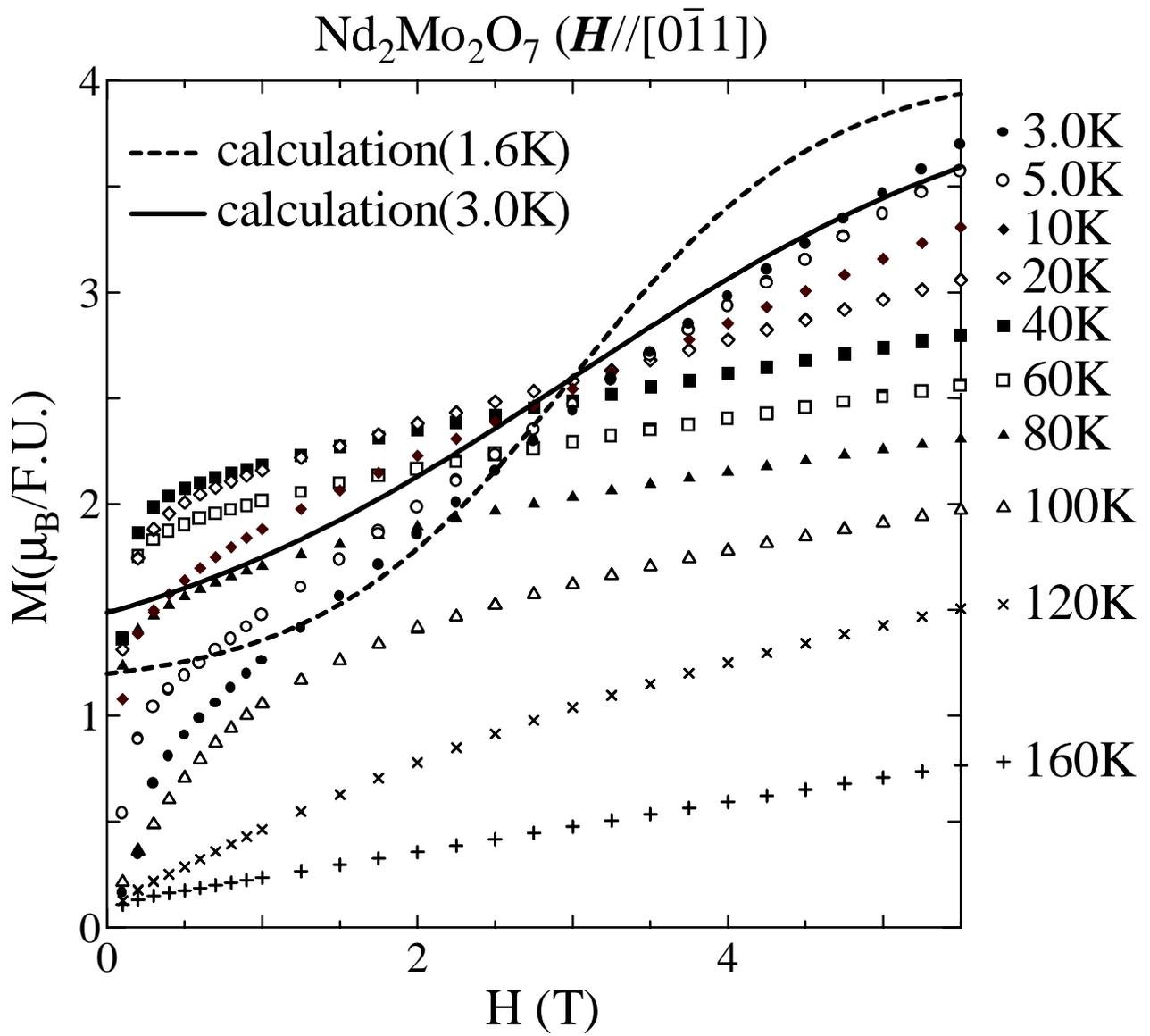

Fig. 6

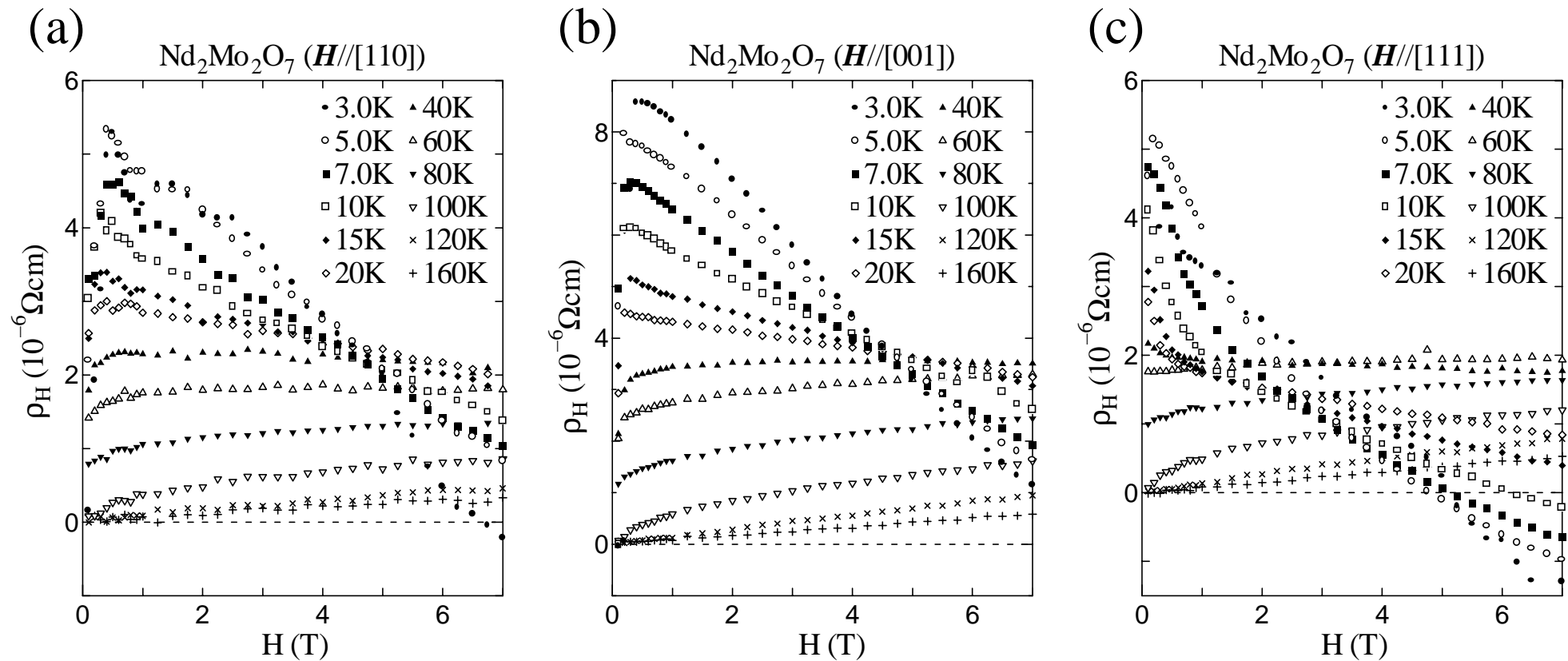

Fig. 7

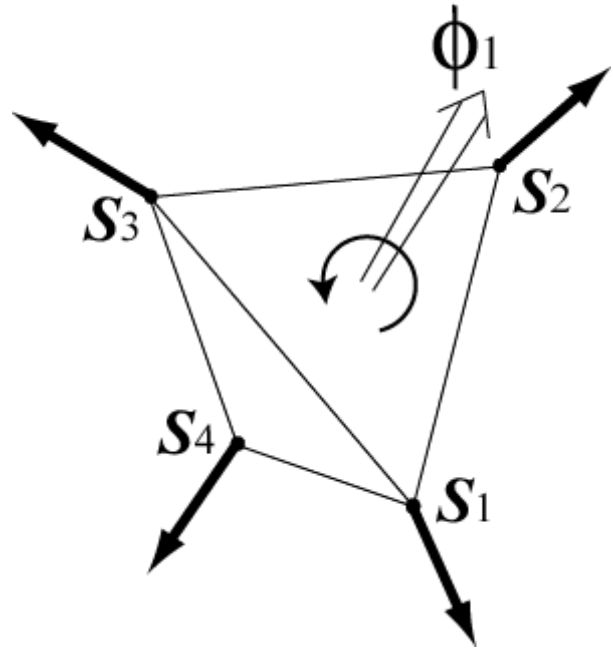

Fig. 8

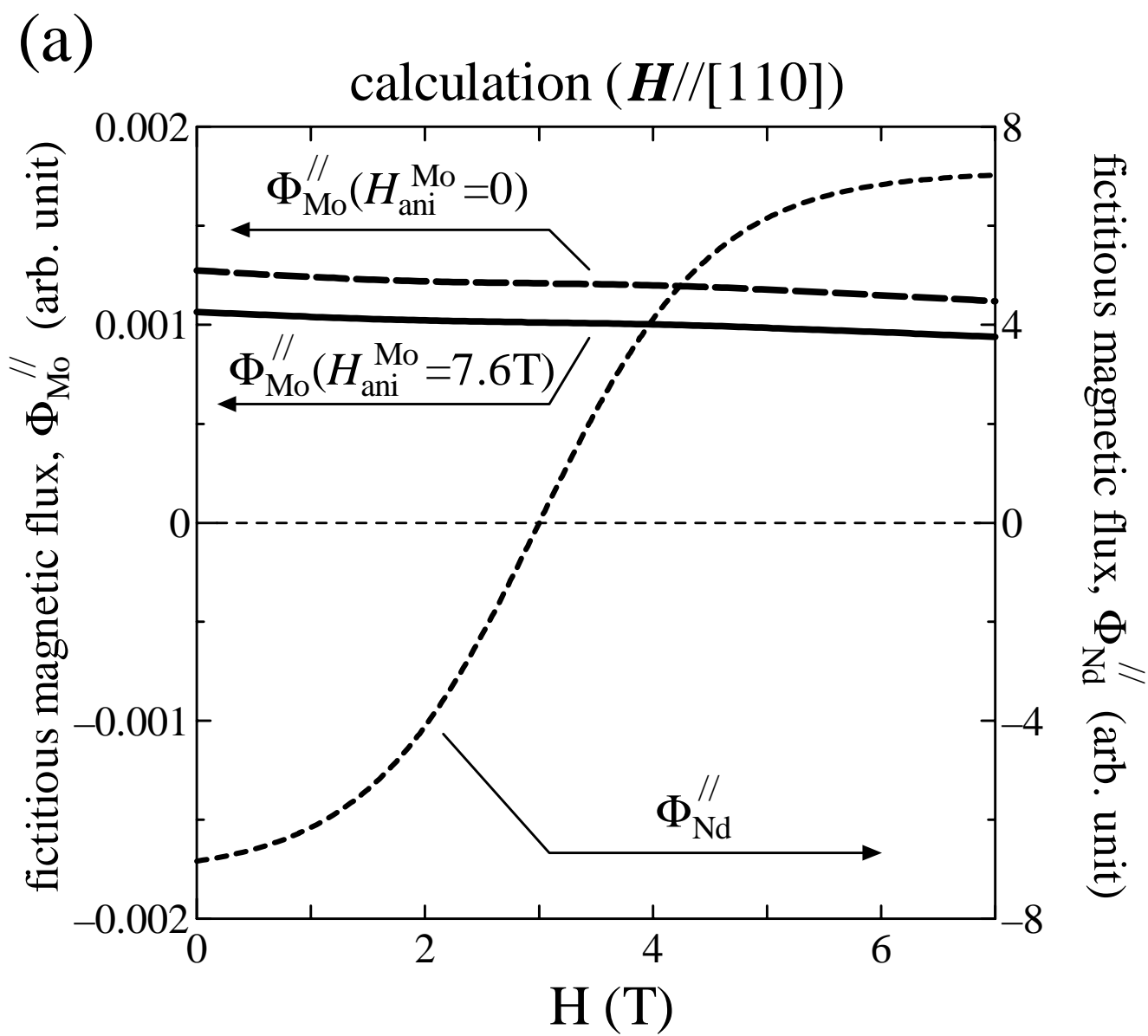

Fig. 9(a)

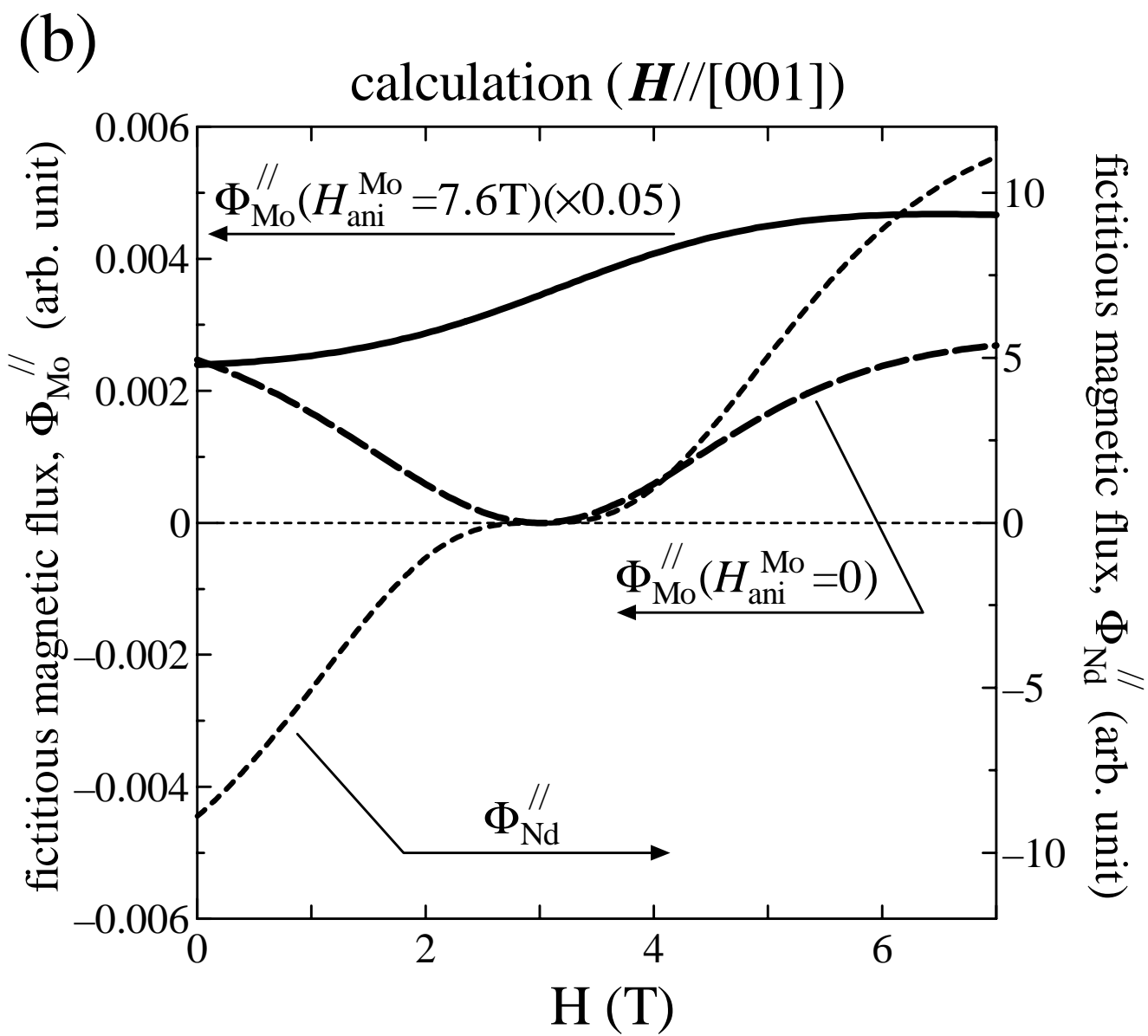

Fig. 9(b)

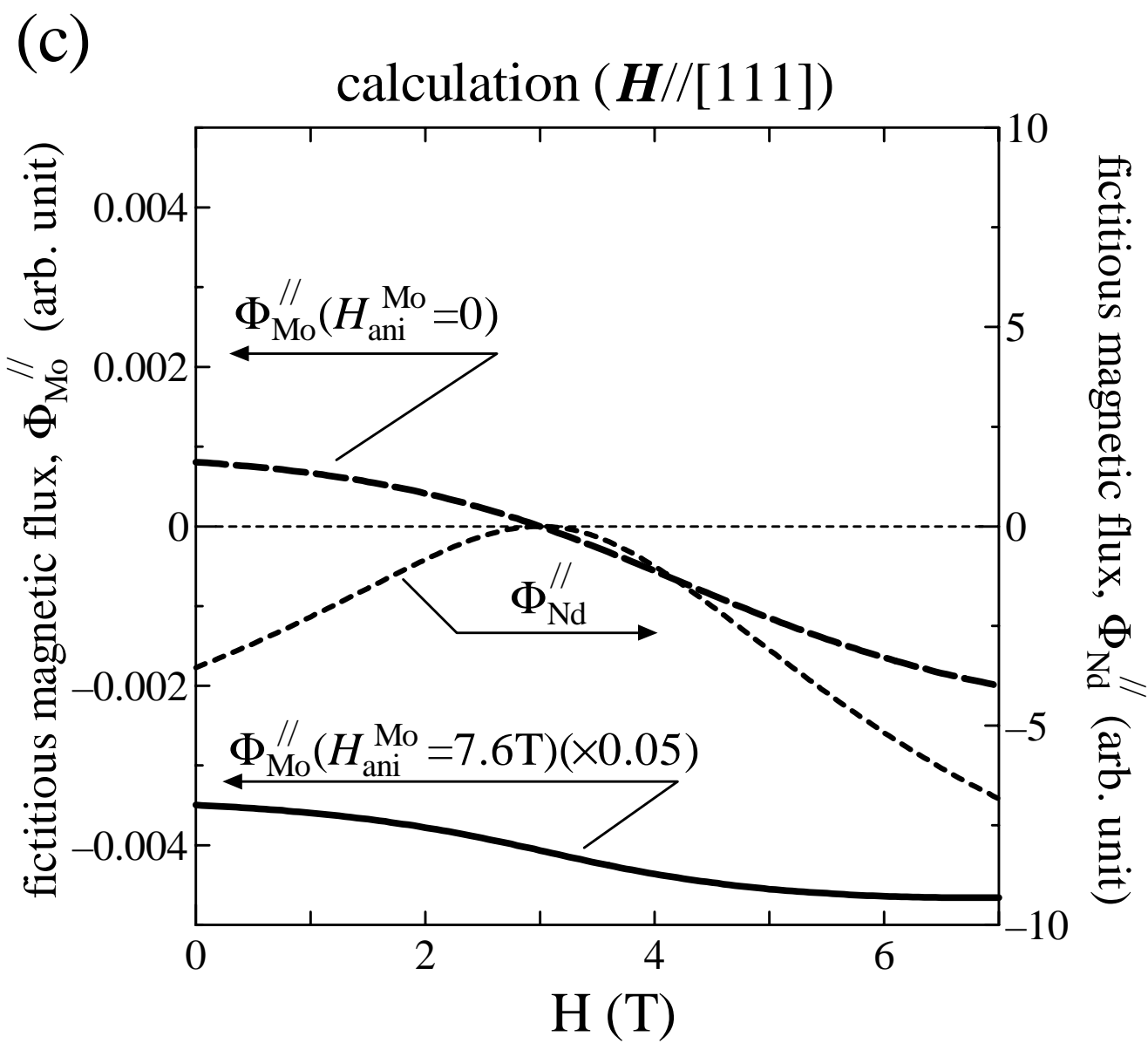

Fig. 9(c)

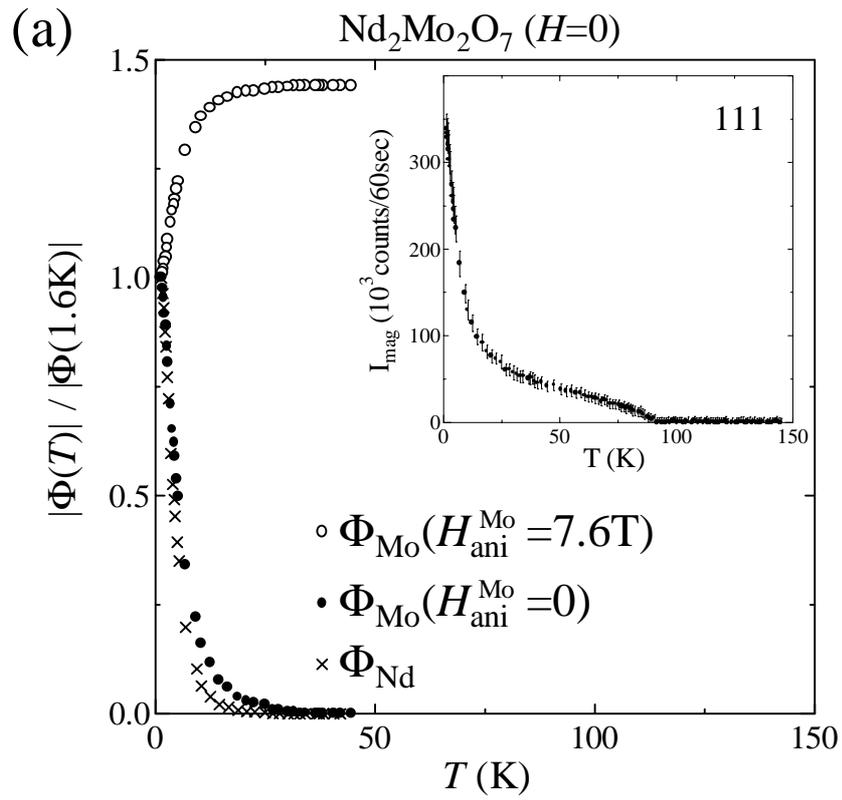
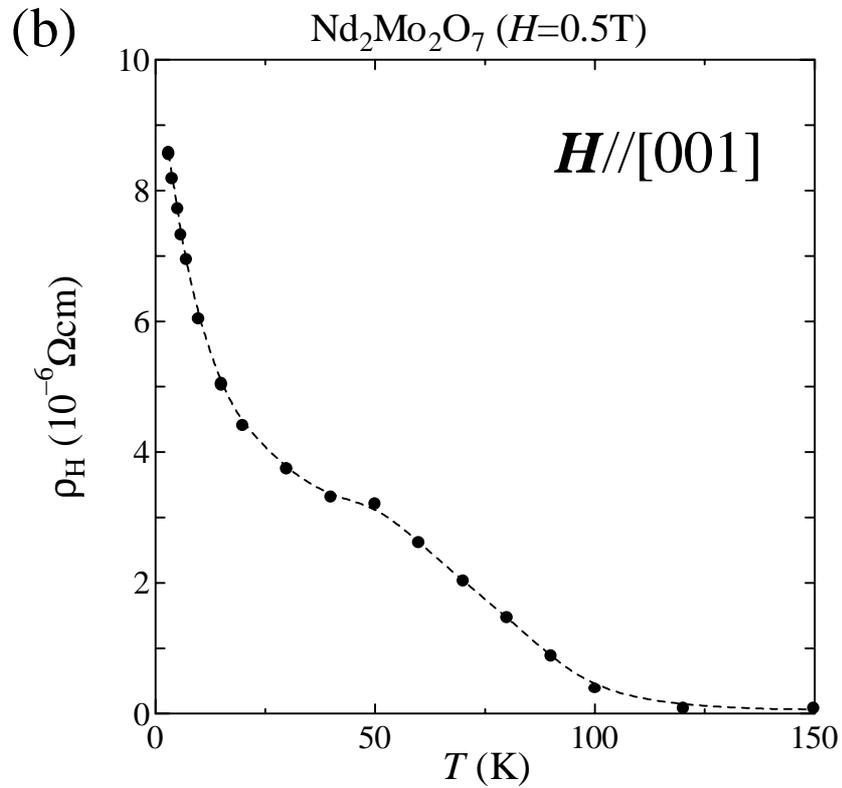

Fig. 10